\documentclass[acmsmall,authorversion]{acmart} 
\AtBeginDocument{%
  }

\begin{document}

\title[Approaching Confidence-Based, Real-Time Line Assignment in Reading Gaze Data]{Sure About That Line? Approaching Confidence-Based, Real-Time Line Assignment in Reading Gaze Data}

\author{Franziska Kaltenberger}
\email{franziska.kaltenberger@tum.de}
\affiliation{%
  \institution{Technical University Munich}
  \city{Munich}
  \country{Germany}
}
\affiliation{%
  \institution{Munich Center for Machine Learning (MCML)}
  \city{Munich}
  \country{Germany}
}

\author{Wei-Ling Chen}
\affiliation{%
  \institution{Technical University Munich}
  \city{Munich}
  \country{Germany}
}

\author{Enkeleda Thaqi}
\affiliation{%
  \institution{Technical University Munich}
  \city{Munich}
  \country{Germany}
}
\affiliation{%
  \institution{Munich Center for Machine Learning (MCML)}
  \city{Munich}
  \country{Germany}
}
\email{enkeleda.thaqi@tum.de}

\author{Enkelejda Kasneci}
\affiliation{%
  \institution{Technical University Munich}
  \city{Munich}
  \country{Germany}
}
\affiliation{%
  \institution{Munich Center for Machine Learning (MCML)}
  \city{Munich}
  \country{Germany}
}
\email{enkeleda.kasneci@tum.de}

\renewcommand{\shortauthors}{Kaltenberger et al.}

\begin{abstract}
Remote and webcam-based eye tracking in multi-line reading suffers from various noise factors and layout ambiguity, precisely where real-time reading support needs reliable, per-fixation line assignment. 
Prior work largely addresses this challenge post hoc or by restricting behavior (e.g., disallowing re-reading), undermining interactive use.  
We propose CONF-LA (Confidence-score-based Online Fixation-to-Line Assignment), a principled, low-latency approach that integrates knowledge about reading behavior and Gaussian line likelihoods over fixations to compute a posterior-line-score and defers assignments when uncertainty is high.
Evaluated on existing open-source data, CONF-LA demonstrates stable performance in post hoc analysis and closes the online-offline gap ($\leq$ 1-2~\%) with a mean per-fixation latency of 0.348~ms. 
Our approach exhibits particular invariance toward regressions, yielding significant improvement in ad hoc median accuracies on children data ($\approx$ 95~\%) over all tested algorithms.  
We encourage further research in this direction and discuss possibilities for future development.
\end{abstract}

%% Rights management information.  This information is sent to you
%% when you complete the rights form.  These commands have SAMPLE
%% values in them; it is your responsibility as an author to replace
%% the commands and values with those provided to you when you
%% complete the rights form.

\setcopyright{cc}
\setcctype{by}
\acmJournal{PACMCGIT}
\acmYear{2026} \acmVolume{9} \acmNumber{2} \acmArticle{19}
\acmMonth{6} \acmDOI{10.1145/3803540}



\begin{teaserfigure}
    \centering
    \includegraphics[width=0.88\linewidth]{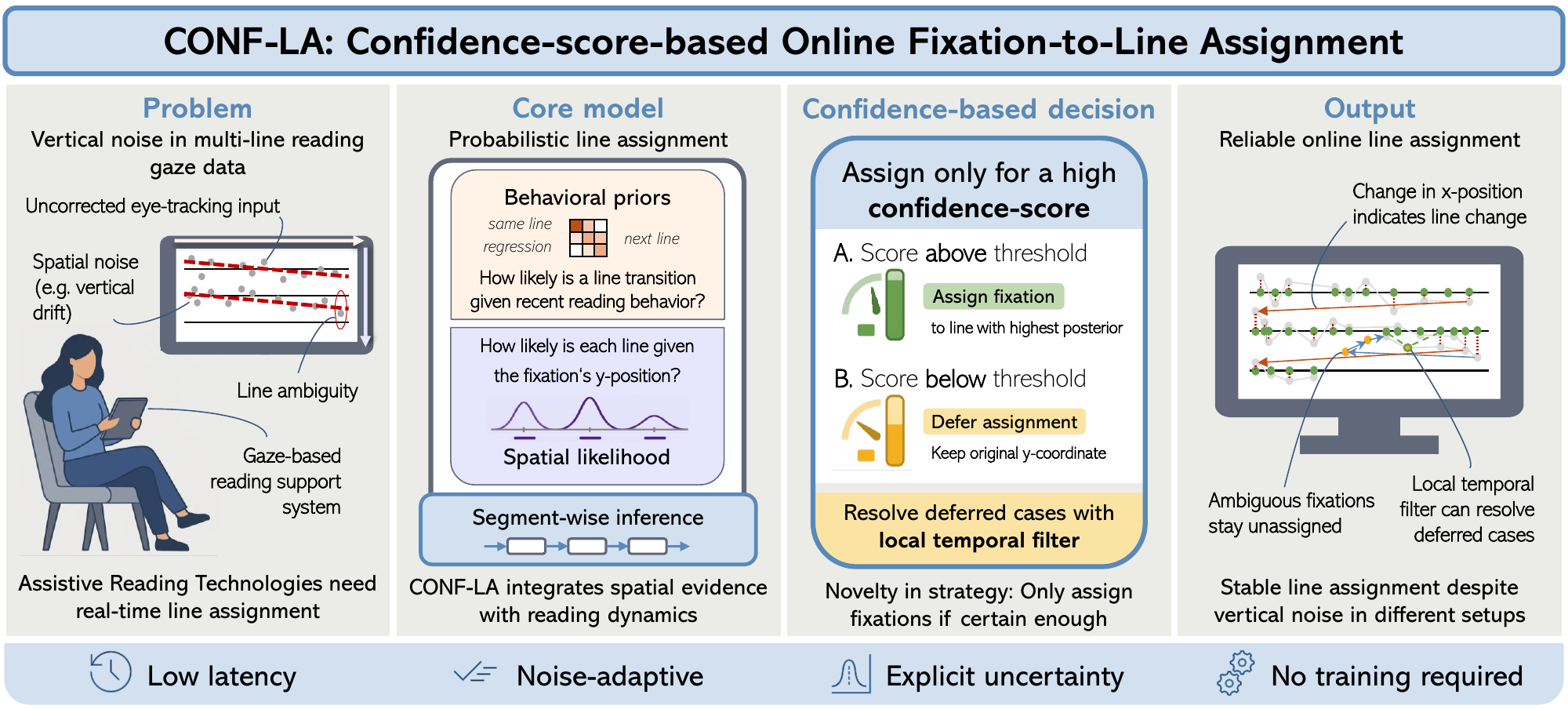}
    \caption{Reliable, real-time assignment for noisy multi-line reading gaze.}
    \label{fig:teaser_image}
    \Description{A four-panel pipeline diagram for CONF-LA showing the transition from noisy gaze data (Problem) through probabilistic modeling (Core Model) and threshold-based filtering (Confidence-based decision) to a stable, assigned output on a monitor.}
\end{teaserfigure}


\maketitle
\section{Introduction}

Eye movements reflect processes in visual information gathering, ranging from large-scale exploration to fine-grained, targeted movements like reading \cite{rayner2009EyeMovementsAttention}.
Reading is defined as the process of extracting information from a sequence of written language, which explicitly extends beyond the identification of isolated words \cite{blythe2014DevelopmentalChangesEye, protopapas2013DevelopmentSerialProcessing}.
This requires integrating information across longer texts \cite{altani2017ContributionExecutiveFunctions, protopapas2013DevelopmentSerialProcessing, vanviersen2025BetweenwordProcessingTextlevel}, 
a challenge, especially for beginning readers who often struggle to process extended passages while maintaining 
attention and spatial orientation \citep{gabrieli2012ReadingAbilitiesImportance, white2019LinkReadingAbility}. 
While educational technologies offer promising reading support (e.g.: \citet{rummens2025GazebasedWordHighlighting, schiavo2021AttentiondrivenReadaloudTechnology, thaqi2024SARASmartAI}), 
these systems often cannot accommodate regressions (e.g., re-reading) or forward jumps, and are highly sensitive to spatial noise in gaze data.  
This is especially problematic for children's eye movements that are inherently more variable and less systematic than those of skilled readers \citep{hindmarsh2021EyeMovementPatterns, reichle2015using}. 

These design choices reflect the challenges of vertical drift and spatial noise in multi-line reading data, which require correction \cite{adedeji2024ChildrensReadingSublexical, reichle2015using, mercier2024DualInputStream, carr2022AlgorithmsAutomatedCorrection}. 
Existing solutions focus on post hoc analysis (e.g., \citet{carr2022AlgorithmsAutomatedCorrection, almadi2024AdvancingDynamicTimeWarp, vadillo2015SimpleAlgorithmOffline}).
More recent advances based on transformers \citep{mercier2024DualInputStream} or CNNs \cite{shangareev2025ReadingProgressTracking} improve accuracy and but still rely on completed reading trials or restrict re-reading, limiting their use in real-time applications.  
Children's reading data, showing more noise and regressions \cite{blythe2011ReadingDisappearingText, rayner2012psychology}, demands especially resilient algorithms. 
Manual correction still outperforms algorithms \cite{almadi2025ValidityBenefitManual} but is not an option in assistive technologies. 

To fill this gap, we present CONF-LA (Confidence-score-based Online Fixation-to-Line Assignment), 
fusing Gaussian line likelihoods with behavioral priors over line changes. 
More specifically, at every fixation, CONF-LA computes a posterior over lines and assigns fixations to a line only when confidence-scores exceeds a formal threshold, deferring otherwise and resolving ambiguities with a short-horizon temporal filter. 
This approach yields three practical benefits: 
(i) supports re-reading without restricting natural reading behavior, 
(ii) explicitly models uncertainty providing interpretability, and 
(iii) is layout-agnostic and plug-and-play, compatible with commodity eye trackers. 
Our analysis establishes CONF-LA as a promising step towards real-time gaze-driven reading.
Building on \citet{carr2022AlgorithmsAutomatedCorrection} and \citet{mercier2024DualInputStream}, we benchmark CONF-LA on simulated and real-world eye-tracking data, demonstrating: 
\begin{enumerate}
    \item \textit{Performance invariance} between ad hoc and post hoc analysis.
    \item \textit{Superior ad hoc accuracy} to all benchmarked methods, especially on noisy and regression-prone data.
    \item \textit{Real-time viability}, with minimal latency and improved accuracy over baseline in simulated real-time. 
    \item \textit{Generalizability without learning} to similar reading setups. 
\end{enumerate}

Beyond accuracy, CONF-LA enables adaptivity and interpretability in gaze alignment, thus opening possibilities for new analysis, such as differentiating reading from non-reading gaze behavior \cite{rayner1996MindlessReadingRevisited}.

\section{Related work}
\subsection{Eye-Tracking-Based Reading Technologies}

Recent improvements in eye-tracking accuracy and accessibility have advanced gaze-informed interfaces, paving the way for gaze-driven reading support tools.
Beginning or neurodivergent readers, in particular, can benefit from assistive technologies that guide attention and maintain engagement \cite{keelor2023ImpactTexttospeechFeatures}.
Systems like gaze-driven word highlighting \cite{rummens2025GazebasedWordHighlighting} and attention-driven read-aloud tools \cite{schiavo2021AttentiondrivenReadaloudTechnology} showcase the potential of gaze-informed interfaces, but struggle with robust, real-time inference of readers' position in the text.

These approaches compensate for inaccuracies in gaze estimation by modifying visual layout rather than modeling gaze uncertainty.
\citet{schiavo2021AttentiondrivenReadaloudTechnology}, for example, extended the bounding boxes around words to accommodate noise in the gaze signal and constrained their system to phrase-based read-aloud units that the recorded fixations have to fall into in consecutive order.
As a result, the tool could not capture regressive eye movements or re-reading behavior.
Similarly, \citet{rummens2025GazebasedWordHighlighting} required children’s fixations to fall within the bounding box of the first word in a paragraph to trigger highlighting, and subsequent fixations could not skip more than one word.
Their method, too, did not support re-reading or discontinuous gaze paths.
Other assistive reading systems employ comparable heuristics, like increased text or line spacing, enlarged bounding boxes, or strict calibration routines \cite{minas2025AdaptiveRealTimeTranslation, morita2025GenAIReadingAugmentingHuman}.
While reducing the impact of tracking inaccuracies, they do not treat gaze data as a noisy, probabilistic signal.

\subsection{Automated Reading Gaze Correction} 

Current gaze-based reading technologies are limited by the absence of stable, real-time line-assignment methods for noisy fixation data.
While systematic investigations of ad hoc mapping remain scarce in the literature \cite{shangareev2025ReadingProgressTracking}, extensive work has focused on post hoc line assignment and correction of recorded reading traces \cite{almadi2024AdvancingDynamicTimeWarp, culemann2024SystematicDriftCorrection, hyrskykari2006UtilizingEyeMovements, mercier2024DualInputStream, shamy2023IdentifyingLinesInterpreting,vadillo2015SimpleAlgorithmOffline}.
Several research tools, including PopEye \cite{schroeder2019popeye}, GazeGenie \cite{mercier2024GazeGenieEnhancingMultiLine}, and Fix8 \cite{almadi2025CombiningAutomationExpertise}, now incorporate these algorithms into standardized analysis pipelines.

\citet{carr2022AlgorithmsAutomatedCorrection} provided a major consolidation of this field, reviewing a broad set of line-assignment strategies and formalizing them into nine algorithms.
The \texttt{compare} algorithm \cite{sanches2015eye, yamaya2017VerticalErrorCorrection} performed so badly that it was excluded from the following analyses.
Along with their newly proposed dynamic time warping (DTW) method, these algorithms were grouped into three main categories: 
(1) Sequential algorithms (
\texttt{warp} \cite{carr2022AlgorithmsAutomatedCorrection}, 
\texttt{segment} \cite{abdulin2015PersonVerificationEye}) 
that rely solely on the temporal order of fixations and words;
(2) Relative-positional algorithms (
\texttt{cluster} \cite{schroeder2019popeye}, 
\texttt{merge} \cite{spakov2019ImprovingPerformanceEye}, 
\texttt{regress} \cite{cohen2013SoftwareAutomaticCorrection}, 
\texttt{stretch} \cite{lohmeier2015experimental}) 
that incorporate partial geometric information about line positions; and
(3) Absolute-positional algorithms (
\texttt{attach} \cite{carr2022AlgorithmsAutomatedCorrection}, 
\texttt{chain} \cite{hyrskykari2006UtilizingEyeMovements, mishra2012heuristic, schroeder2019popeye}, 
\texttt {split} \cite{carr2022AlgorithmsAutomatedCorrection}) 
that depend primarily on spatial distance to fixed line coordinates.
The latter “quite conservative” \citep[pp.~305]{carr2022AlgorithmsAutomatedCorrection} group introduces minimal corrective bias, but shows limited invariance to non-regressive distortion types, such as drift or slope.
The attach algorithm, which simply assigns each fixation to the closest line, has since become a de facto baseline, and its heuristic logic underlies many practical implementations used in assistive reading technologies.
Thus, \citet{carr2022AlgorithmsAutomatedCorrection} established a shared foundation for evaluating and extending line-assignment methods, enabling reproducible benchmarking and systematic improvement.
Their open data and code resources have since been used by subsequent studies \citep{almadi2024AdvancingDynamicTimeWarp, mercier2024DualInputStream} to test algorithmic stability under controlled distortions and to explore heuristic, hybrid or machine-learning-based correction schemes.

Beyond these deterministic correction algorithms, other studies have explored probabilistic formulations of gaze-text alignment.
\citet{bottos2019TrackingProgressionReading, bottos2019NovelSlipKalmanFilter} proposed a hybrid approach combining a Kalman filter for denoising raw gaze data with a discrete Hidden Markov Model (HMM) for line identification.
While effective in smoothing trajectories, these models assume line-by-line reading progression with fixed transition probabilities and rely on full EM-based parameter training, making them less suited for ad hoc or context-dependent adaptation.
Our approach, by contrast, retains the probabilistic treatment of uncertainty in line assignments but replaces global training with context-sensitive, time-varying transition updates, enabling real-time inference under non-stationary reading dynamics.

\section{Theory} \label{sec:theory} 

CONF-LA is based on the principle of only correcting $y$-values if the certainty is high enough to ensure that this correction yields the correct line assignment.
We first present the general algorithmic principle of this approach and then provide detailed information about our implementation of various components. 
Please note that the principle is explicitly formalized to allow for different implementations of specific steps to be explored in future research.

\subsection{CONF-LA: Confidence-score-based Online Fixation-to-Line Assignment} \label{sec:principle}
We present the general algorithmic principle for a confidence-score-based line assignment for online fixation correction. 
It assumes $N$ fixations recorded from participants whilst reading a text with $M$ lines of a maximum line length $L$. 
We divide fixations into segments $\mathbf{S}$ with fixation 
\begin{math}
    (y_{i}^{(s)}, x_{i}^{(s)}), i=1, ..., S
\end{math} being the $i$-th fixation recorded in the segment. 

\begin{enumerate}
    \item For every fixation, get the \textit{\textbf{ad hoc line distance likelihood}} \begin{math}
        (d_{l})_i^M = \mathbf{d}_i
    \end{math} 
    for each line $l\leq M$. 
    \item Analyze the fixation data for information related to the \textit{\textbf{reading context}} to modify a \textit{\textbf{behavioral prior}}. 
    \item Integrate this information over the segment to compute an \textit{\textbf{assignment posterior}} $\mathbf{A}^{(s)}$ over lines to obtain 
    \begin{enumerate}
        \item \textit{\textbf{line assignments}} for all fixations in the segment: 
        \begin{math}
            \mathbf{a}^{(s)}_{i} = \arg \max_{l \leq M}~\mathbf{A}^{(s)}_{i,l}    
        \end{math}
        \item \textit{\textbf{confidence-score values}} for line assignments: 
        \begin{math}
            \mathbf{c}^{(s)}_{i} = \max_{l \leq M}~\mathbf{A}^{(s)}_{i,l}
        \end{math}
    \end{enumerate}
    \item A fixation $i$ only gets assigned to a line ($y_{i} = y_{l}$ for $l=a_{i}$) if its confidence-score value has reached a set \textit{\textbf{confidence threshold}} $C$ (i.e. $c_{i} \geq C$). 
    \item For fixations with $c_i < C$, \textit{\textbf{neighboring line assignments}} can update the confidence-score value $c_i$ to potentially exceed the confidence threshold or to indicate a different line $l' \neq a_i$ with a confidence $c_i'$. 
    If those corrected confidence-score values reach the threshold $C$ (i.e. $c'_{i} \geq C$), a fixation's $y$-value gets updated accordingly.
    \item Otherwise, fixation coordinates stay \textit{\textbf{unchanged}}. 

\end{enumerate}

\subsection{HMM-CONF-LA: Generative Model and Segment-Based Inference} \label{sec:gen_mod}

\begin{figure}
    \centering
    \includegraphics[width=0.95\linewidth]{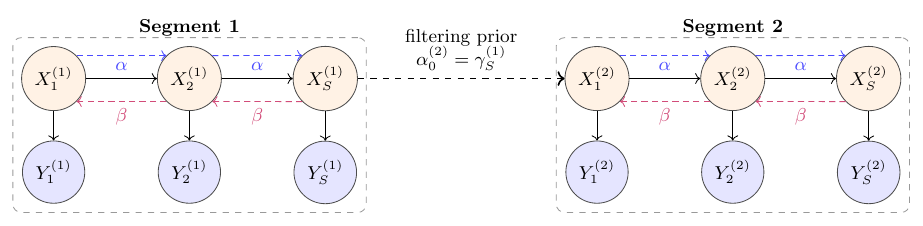}
    \caption{Visualization of the generative model and inference procedure used to obtain confidence values. Hidden line states $X$ are changed with transition probabilities provided by $\mathbf{L}$ and emit normally distributed observed y-positions of fixations. During inference, we perform global filtering ($\alpha$) across, and local smoothing ($\beta$) within segments of size $S$.}
    \label{fig:generative_model_inference}
    \Description{
    A probabilistic graphical model diagram illustrating inference over time across two segments. Segments are given as two identical, rectangular dashed boxes, each containing a part of the Hidden Markov Model. The top-row, orange hidden state nodes, $X_1$ to $X_S$, are linked sequentially and vertical arrows pointing down to bottom-row, purple observed fixation nodes, $Y_1$ to $Y_S$.

    Inference Flow: Blue solid arrows labeled with alpha run horizontally between hidden states, showing forward filtering. Red dashed arrows labeled with beta run horizontally backward between hidden states, showing backward smoothing within a segment.

    Segment Link: A dashed horizontal line connects the last hidden state of Segment 1 to the first hidden state of Segment 2. Text above this line defines the filtering prior for Segment 2 as being equal to the smoothed posterior from the last timestep of Segment 1.}
\end{figure}

We model the relation between lines and gaze position as a non-homogeneous Hidden Markov Model (HMM) where hidden states $X_i$ represent discrete lines and observations $Y_i$ correspond to the continuous $y$-position of a fixation $i$. 
The generative model is given by 
(i) a uniform prior over all lines,
(ii) a continuous Gaussian emission probability distribution, and
(iii) a discrete, context-sensitive transition probability distribution. 
Appendix \ref{apdx:gen_mod} provides detailed mathematical specification.
We assume sequential processing of segments containing $S$ fixations each to support ad hoc assignment (i.e. $S<N$).
Inference is performed using segment-wise forward-backward message passing to compute posterior confidence scores within a segment.
Appendix \ref{apdx:equations} details the exact inference equations.
For real-time line assignment ($S=1$), inference becomes a mere filtering procedure. 
For post hoc line assignment ($S=N$), inference becomes a full smoothing procedure. 
Figure \ref{fig:generative_model_inference} visualizes the generative model and the inference procedure over segments.

\subsection{Line-distance Likelihood with Adaptive Shift} \label{sec:adaptive_shift}
We assume ad hoc line distance likelihoods to be normally distributed \cite{niehorster2020CharacterizingGazePosition, pekkanen2017NewGeneralApproach}.
The standard deviation is a proportion of the line distance $d_l$, remaining constant over lines: $\sigma = d_l \cdot \sigma_{scale}$ ($\sigma_{scale} \in \mathbb{R}^+$).
Higher $\sigma_{scale}$ models more divergence from the line.
The mean is the proportionally shifted center of the current line $y_{l}$: $\mu_l = y_{l} \cdot \mu_{shift}$ ($\mu_{shift} \in \mathbb{R}^+$). 
For $\mu_{shift} < 1.0$, the upward shift increases for later lines, and vice versa.

For $S>1$, we included the option to use an adaptive shift $\mu_{shift}$ for the Gaussian line likelihood.  
If the adaptive shift is used, we initialize $\mu_{shift}=1$.
After correcting the line assignments in a segment, the shift gets adapted if a majority of fixations have been corrected to the line above or below (i.e. $\mu_{shift}=0.99$ or $\mu_{shift}=1.01$, respectively). 
We motivate this adaptivity in the following Section and explicitly state when the adaptivity was used. 

\subsection{Context and transition matrix} \label{sec:contextAtransition}

Behavioral priors for skilled reading are encoded in a transition matrix $\mathbf{L}$, modeling the probability of a fixation $(y_{i}, x_{i})$ originating from a different line than the previous fixation $(y_{i-1}, x_{i-1})$. 
The matrix is line-agnostic, assigning identical values to all lines $l\leq M$. 
These priors and updates encode assumptions on typical skilled reading behavior and are not learned from data.
Most fixations in a skilled reader's scan path occur during continuous forward in-line reading ($\mathbf{L}_{l,l} = 0.85$),  
most likely progressing to the following line ($\mathbf{L}_{l,l+1} = 0.1$), but occasionally re-reading of the previous line ($\mathbf{L}_{l+1,l} = 0.03$). 
The matrix updates dynamically based on saccade velocity and sudden x-position changes, reflecting deviations from forward in-line reading.
Finally, the context-aware transition matrix is normalized ($\sum \mathbf{L}_{l, :} = 1.0$). 

\subsubsection{Return Sweeps} 
According to \citet{rayner1998eye}, saccadic peak velocity is a monotonic function of saccade amplitude, i.e., larger eye movements achieve a significantly higher speed. 
While within-line saccades are relatively short (2~° of visual angle in 30~ms \cite{rayner2009EyeMovementsAttention}), return sweeps must traverse almost the entire horizontal width of the line \cite{slattery2019EyemovementExplorationReturnsweep}. 
In laboratory settings, these movements can reach velocities of up to 500~°/s \cite{rayner1998eye}. 
As obtaining angular velocities is difficult in non-laboratory settings due to missing distance information, we use a conservative\footnote{
Reading saccade: $\sim$\,2~° in 30~ms $\rightarrow v=67$~°/s. Non-reading threshold (distance: 100~cm): $v=300$~cm/s $\approx$ 172~°/s$~ > 2.5~ \cdot ~$67~°/s
}, linear threshold. 
When the (estimated)\footnote{Ideally, saccade velocity is accessible. Otherwise, the algorithm can use an estimated peak velocity (see Equation \ref{eq:velocity_estimate}).} peak velocity exceeds a set threshold (
\begin{math}
    \hat{v}_i > 300 \scriptstyle \frac{cm}{s}
\end{math}), the context should favor a line change. 
This is reflected in the transition matrix by setting $\mathbf{L}_{l,l} = 0.05$ and $\mathbf{L}_{l,l+1} = 0.9$, respectively.

\subsubsection{Sudden changes in the x-position} 
A line change in skilled reading typically starts 5-7 characters before the end of a line and lands 3-7 characters into the next line \cite{rayner2009EyeMovementsAttention, slattery2019EyemovementExplorationReturnsweep}. 
As reliable character-level tracking is infeasible in a non-laboratory setting, we use a percentage of a line length to obtain a generalizable, simplified proxy.
\citet{rayner1998eye} explicitly notes that because readers do not typically fixate at the margins of a line, "about 80~\% of the text typically falls between the extreme fixations" \citep[pp.~375]{rayner1998eye}.
Thus, we favor a line transition if the change in x-position between subsequent fixations covers more than 80\% of the line length $l_{length}$ (i.e. $x_{i-1} -x_{i} < -0.8 \cdot l_{length}$). 
This is reflected in the transition matrix by setting $\mathbf{L}_{l,l} = 0.3$ and $\mathbf{L}_{l,l+1} = 0.7$, respectively.

\subsection{Neighbor Voting}
For $S>1$, we apply a window correction to unassigned fixations, using information about the $W$ neighboring fixations before and after the unassigned fixation $i$. 
If fixation $i$ remained unassigned to line $\overline{l}$ (\begin{math}
    c_{\overline{l}} < C
\end{math}), but a percentage of neighbors agree on line $\hat{l}$ with $c_{\hat{l}} > \Omega$
and line $\hat{l} \neq \overline{l}$, fixation $i$ is assigned to line $\hat{l}$.
If $\hat{l} = \overline{l}$, the fixation's line assignment only gets updated if 
\begin{math}
    c_{\overline{l}} (1-\omega) + c_{\hat{l}} \omega > C
\end{math}.
We set $W=5$ and $\omega = 0.4$.

\section{Methods} \label{sec:analysis}
We performed various tests on our algorithm to 
(1) analyze parameter effects, including segment size, 
(2) compare it to algorithms benchmarked by \citet{carr2022AlgorithmsAutomatedCorrection}, 
(3) investigate algorithmic components,
and (4) test generalization to other datasets without fine-tuning parameters.
Building on the implementation of \citet{carr2022AlgorithmsAutomatedCorrection}, we provide open-source code\footnote{Open source implementation and data available at \url{https://gitlab.lrz.de/hctl/conf-la}} to reproduce our findings and use a streamlined CONF-LA implementation in reading technologies. 

\subsection{Data}
\subsubsection{Datasets}
We extended the analysis by \citet{carr2022AlgorithmsAutomatedCorrection} using their fixation data, a 5.5~\% subset from \citet{pescumaeyereadit}, containing 24 silent reading trials for 8- to 11-year-old children and 24 trials for adults across 12 passages ($\sim$130 words, 10-13 lines).
Text was displayed in 20~pt. Courier New with 64 pixels line height and 80 characters per line. 
Eye movements were recorded from the right eye using an EyeLink 1000 Plus (1000~Hz). 
Data was cleaned by removing fixations outside text bounds or occurring before and after reading. 
Two correctors manually assigned lines, creating a gold standard. 
Disagreements between correctors were more frequent for discarded fixations than for assigned lines. 
For generalization testing, we used the MECOde dataset \cite{siegelman2022ExpandingHorizonsCrosslinguistic}, processed by \citet{mercier2024DualInputStream}, which yielded the lowest optimal accuracy.
This provided additional 648 adult reading trials on passages of 10-14 lines. 

\subsubsection{Velocity estimation}
Unlike other algorithms, CONF-LA requires saccade information for context extraction. 
Online recognition of fixations and saccades can be effectively achieved with probabilistic methods \cite{kasneci2014applicability,kasneci2015online}.
We estimated mean saccade velocity as
\begin{math} \label{eq:velocity_estimate}
    \overline{v}_i = {d_{screen}}/({t^{start}_{i+1} - t^{end}_i}) ~   \frac{cm}{s}
\end{math},
where $d_{screen}$ is the on-screen distance between consecutive fixations, assuming \begin{math}
    t^{end}_i < t^{start}_{i+1}
\end{math}. 
Peak saccade velocity was estimated as 
\begin{math}
    \hat{\mathbf{v}}_i = \overline{\mathbf{v}}_i \cdot 1.5
\end{math}.

\subsubsection{Simulation}
Using the framework by \citet{carr2022AlgorithmsAutomatedCorrection}, we tested distortions in reading gaze data, i.e., general noise, shift, slope, within-line, and between-line regressions, as well as additional simulations of line length and line height. 
We further extended the analysis by also assigning simulated fixations in an ad hoc manner for $S=10$. 
For each distortion parameter setting, 20 fixation sequences were simulated. 
Fixation simulations ran on one randomly selected passage from the 12 passages used in the real-world experiment. 
Regression probabilities ($p_{within}$ and $p_{between}$) ranged between 0 (never regress) and 1 (within: regress after every normal fixation; between: regress to any previous line once per line).
Simulated fixations $(y'_i, x'_i)$ on line $l$ included random $x$ within-word location and a distorted $y$ location\footnote{Adapted from Equation (2) in \citet{carr2022AlgorithmsAutomatedCorrection}.}: 
\begin{equation}\label{eq:y_distortion_sim}
    y'_i = \mathcal{N}(y_l, d_{noise}) + x'_i d_{slope} + y_l d_{shift} 
\end{equation}
with $d_{noise} \in [0;40]$ for general noise, $d_{slope} \in [-0.10; 0.10]$ for upward ($<0$) or downward ($>0$) movement within a line, and $d_{shift} \in [-0.20;0.20]$ for increasing upward ($<0$) or downward ($>0$) shift between lines. 
20 values were tested for regression distortions, 40 values for other distortions. 
If not currently tested, parameters were set to have no effect.
For text parameter testing, we used dummy passages with added noise ($d_{noise} = 15$).
Line length $l_{length}$ was varied from 15 to 110 pixels (20 tested values).
Line height $l_{height}$ was tested for every other value ranging from 28 to 104 pixels (39 tested values).
No-effect text parameters correspond to the original passage settings. 

\subsubsection{Augmenting low-fidelity data}
The widespread adoption of eye-tracking-based reading technologies is currently hindered by the limited availability of high-precision hardware \cite{angele2025HowLowCana,niehorster2020impact}.
Scalable line assignment algorithms must therefore also perform well on low-frequency (30-60~Hz), spatially distorted reading data. 
We simulated missing and inaccurate fixation information obtained from low-quality recording conditions using an adapted "drop" approach \cite{angele2025HowLowCana}, applying three degradation steps (see Appendix \ref{apdx:degrad}): 
(i) duration decimation: removing all fixations shorter than one simulation clock time step, set to 33.3~ms (30~Hz). 
(ii) temporal binning: aligning fixation start and end time to the nearest sampling clock time. 
(iii) spatial noise: adding additional spatial distortion to the original $y$ coordinate following Equation \ref{eq:y_distortion_sim}, setting parameters as follows: $d_{noise}=10$, $d_{slope}=0.02$, $d_{shift}=0.02$.
This method preserves original manual corrections on an existing dataset as gold standard.

\subsection{Performance measures}
Performance was evaluated using the percentage accuracy of line assignment compared to the gold standard.
Unassigned fixations were mapped to the closest line for comparability to algorithms that always assign a line to a fixation. 
In reading technologies, this fixation would still fall into an area of interest surrounding a word. 
We emphasize that this accuracy is designed to evaluate the performance in the context of assistive reading technologies. 
Different metrics and datasets are required to evaluate, for example, whether non-reading fixations are correctly left unassigned. 

\subsection{Parameter choice and analysis design}
We performed a grid search to optimize parameter combinations and evaluate their effects on performance. 
Appendix \ref{apdx:grid_search} details grid ranges and results. 
Other parameter settings (e.g. context threshold, transition matrix entries) remained as detailed in Section \ref{sec:theory}.
Optimal parameter settings were identified for various segment sizes ($S=N, 10, 4, 3, 2, 1$). 
We used optimal values for post hoc analysis ($S=N$) in the segment size analysis, the ablation analysis, and the confidence calibration analysis. 
Segment size analysis also included optimal values for $S \leq 5$ to compare full potential with suboptimal parameter choices. 
Simulation, runtime, degradation, and generalization analysis use optimal values for the respective segment sizes optimized for the Carr dataset. 
The first parameter analysis varies the range of confidence threshold values or Gaussian values ($\mu_{shift}$ and $\sigma_{scale}$), respectively, using optimal parameters elsewhere.
Subsequent analyses use the adaptive shift.

\section{Analysis results}

\subsection{Parameter Analysis} \label{sec:P_analysis}

\begin{figure}
  \centering
  \includegraphics[width=\linewidth]{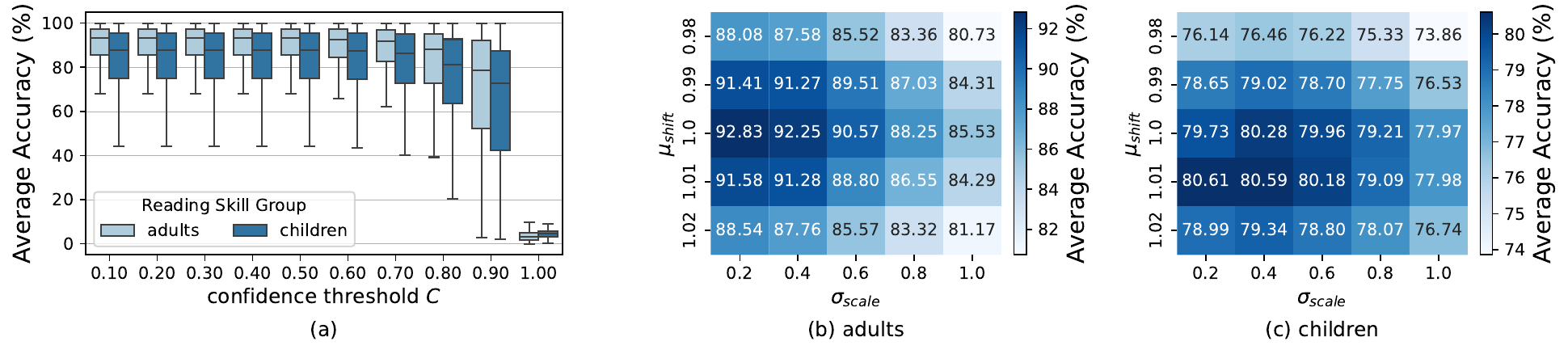}
  \caption{Grid search results averaged over segment sizes show impact of values for (a) confidence threshold (whiskers: $1.5\ IQR$) and (b, c) Gaussian line likelihood parameters. Results indicate that greater uncertainty enables more accurate line assignment. Adults show a more stable reading gaze pattern, whereas children's gaze data is more shifted and displays a wider variation of noise.   
  }
  \label{fig:boxplots_confThresh_MeanVar}
  \Description{
  Three plots (a, b, and c) showing grid search results for accuracy.
    Sub-figure (a): A box plot comparing average accuracy (y-axis) against confidence threshold C (x-axis) from 0.10 to 1.00. Light blue bars represent adults and dark blue bars represent children. Accuracy remains stable and high for both groups until threshold 0.80, after which accuracy for children drops significantly and becomes more volatile, ending near zero at threshold 1.00.
    Sub-figure (b) and (c): Two heatmaps showing accuracy for adults and children based on Gaussian parameters: sigma-scale (x-axis) and mu-shift (y-axis). Both heatmaps use a blue color scale where darker indicates higher accuracy.
    Trends: In both heatmaps, the highest accuracy (darkest blue) is clustered in the center-left. Adult accuracy (b) peaks at 92.83\%, while children's accuracy (c) peaks lower at 80.61\% and shows a more dispersed distribution of high-performance values.}
\end{figure}

Grid search analysis reveals how individual parameters affect algorithmic accuracy when averaged across all remaining parameters, including segment size.
Figure \ref{fig:boxplots_confThresh_MeanVar}(a) shows that allowing for more uncertainty ($C<0.9$) before correcting a line is crucial for correct line assignment accuracy in both children and adult reading gaze data. 
This validates our confidence-score-based approach. 
Consequently, we restrict subsequent analyses to $C<0.9$.

Our approach, by design, also allows insight into reading behavior. 
In particular, optimal Gaussian line likelihood parameters reveal differences between skilled and beginning readers. 
Figure \ref{fig:boxplots_confThresh_MeanVar}(b,c) compares the averaged grid search accuracies. 
Besides generally higher accuracies, adults (b) exhibit more stable reading behavior reflected in less variance in both dimensions, as well as peak accuracy at minimal drift ($\mu_{shift}=1.0$) and minimal noise ($\sigma_{scale} = 0.2$). 
Children (c) show more variability in noise ($\sigma_{scale}$), indicating less consistent gaze data, and increased vertical drift reflected in peak accuracies at $\mu_{shift}=1.01$.
To address the $\mu_{shift}$ differences between beginning and skilled readers, we introduced the adaptive shift of the mean presented in Section \ref{sec:adaptive_shift}.

\subsection{Simulation Analysis} \label{sec:simulation_analysis}
\begin{figure}
  \centering
  \includegraphics[width=0.91\linewidth]{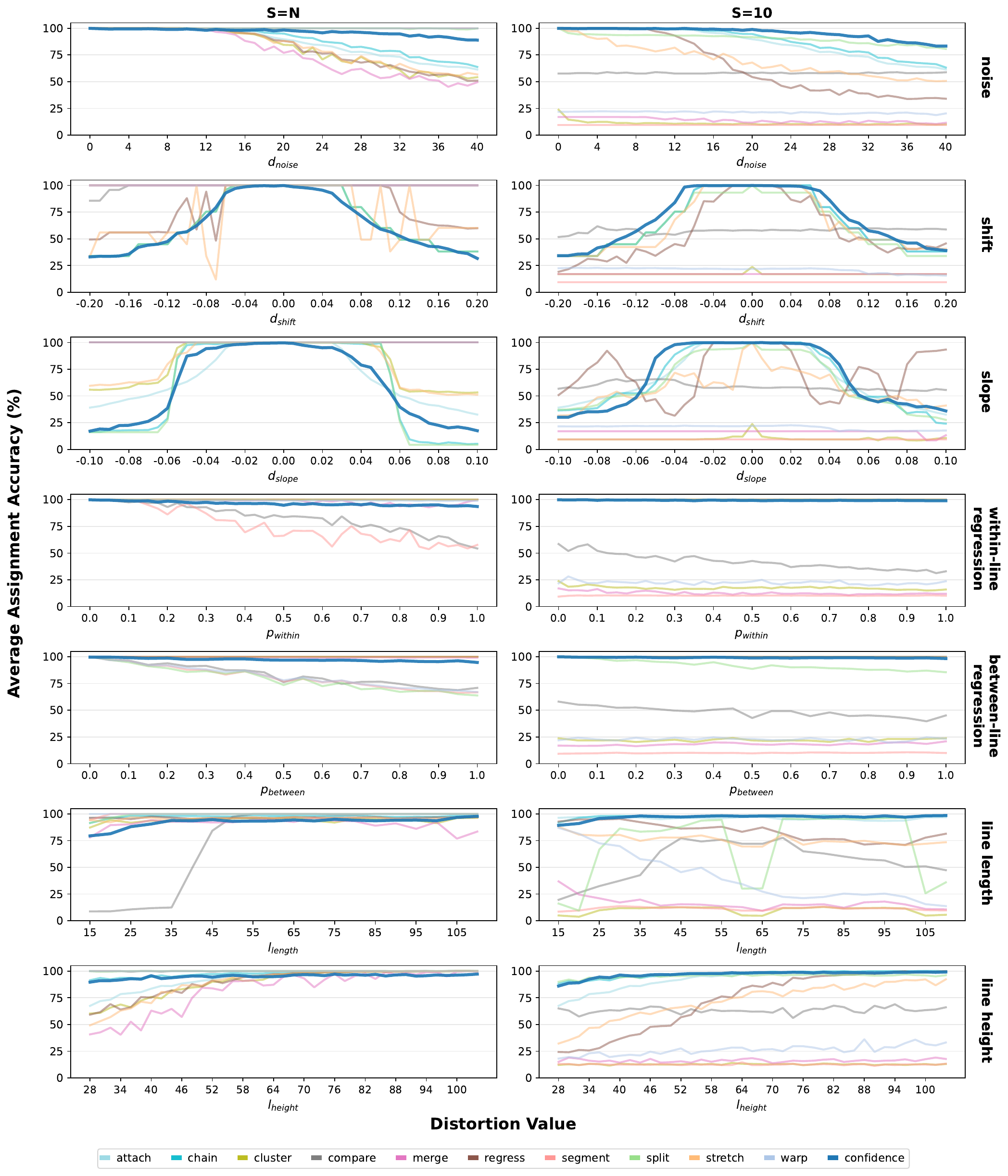}
  \caption{Simulation analysis to test the algorithm's susceptibility to different types of distortion: general noise, shift, slope, within and between line regressions. Average assignment accuracies are compared between post hoc ($S=N$, left) and ad hoc  assignment ($S=10$, right). While all other algorithms show a drop in performance between segment sizes, our confidence approach (thick blue line) is almost invariant to segment size. }
  \label{fig:simulation_analysis}
  \Description{A grid of 14 line graphs arranged in two columns (S=N and S=10) and seven rows representing different distortion types: noise, shift, slope, within-line regression, between-line regression, line length, and line height. Each graph plots Average Assignment Accuracy (0 to 100\%) against increasing distortion values. A thick, dark blue line representing the "confidence" method consistently maintains high accuracy across almost all plots and both columns. Only for high degrees of shift and slope this method sees a drop in performance. In contrast, the other 11 baseline algorithms (represented by thinner, multi-colored lines) show varying degrees of performance degradation as distortion increases, particularly in the S=10 (ad hoc) column.}
\end{figure}

Our simulation analysis compares our confidence-score-based approach to existing benchmarks, focusing on performance across segment sizes. 
Rows in Figure \ref{fig:simulation_analysis} present the results for seven types of distortion: (1) general noise, (2, 3) vertical drift and slope, (4, 5) within/between-line regressions, and (6, 7) line length and height. 
Columns show average assignment accuracies for post hoc correction ($S=N$) and ad hoc correction ($S=10$). 

CONF-LA exhibits minimal sensitivity to segment sizes, as performance remains stable across distortions. 
This segment invariance distinguishes it from most competing algorithms, which show substantial accuracy losses in ad hoc assignment.
Assignment accuracy remains high for general noise ($N: ~> 88~\%, 10: ~>83~\%$), reflecting the model's ability to capture vertical noise. 
However, these results allow only limited comparison with real-world reading data, as the simulation introduces Gaussian noise, which serves as the model's emission distribution.
Furthermore, CONF-LA is particularly stable under regressions (\textit{within-line}: $N: ~> 93~\%, 10: ~>98~\%$; \textit{between-line}: $N: ~> 94~\%, 10: ~>98~\%$), indicating that the transition matrix effectively captures re-reading structure. 
It shows some susceptibility, especially in post hoc analysis, to short lines ($N: ~> 80~\%, 10: ~>92~\%$) and small line heights ($N: ~> 79~\%, 10: ~>77~\%$). 
In contrast, accuracy declines sharply for high degrees of slope ($N: ~> 17~\%, 10: ~>30~\%$) and drift ($N: ~> 31~\%, 10: ~>34~\%$).
These distortions represent the primary limitation of CONF-LA, likely stemming from restricted emission adaptivity.

Compared to the other tested algorithms, CONF-LA maintains higher accuracies in ad hoc assignment across most distortion types, but hardly provides advantages under extreme slope and drift distortion.
Since competing methods rely on sequential information, they exhibit strong performance in post hoc assignment but show decreased accuracy when the segment size is reduced. 
Overall, simulation results indicate that combining spatial likelihood information with behavioral transition assumptions enables stable ad hoc assignments under moderate distortions. 

\subsection{Real-Time Analysis} \label{sec:real_time_analysis}

\begin{figure}
  \centering
  \includegraphics[width=\linewidth]{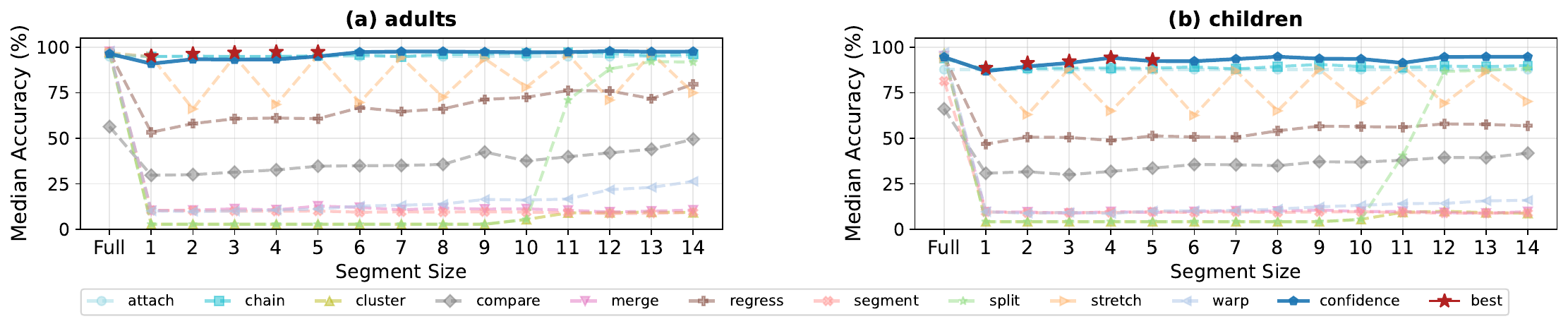}
  \includegraphics[width=\linewidth]{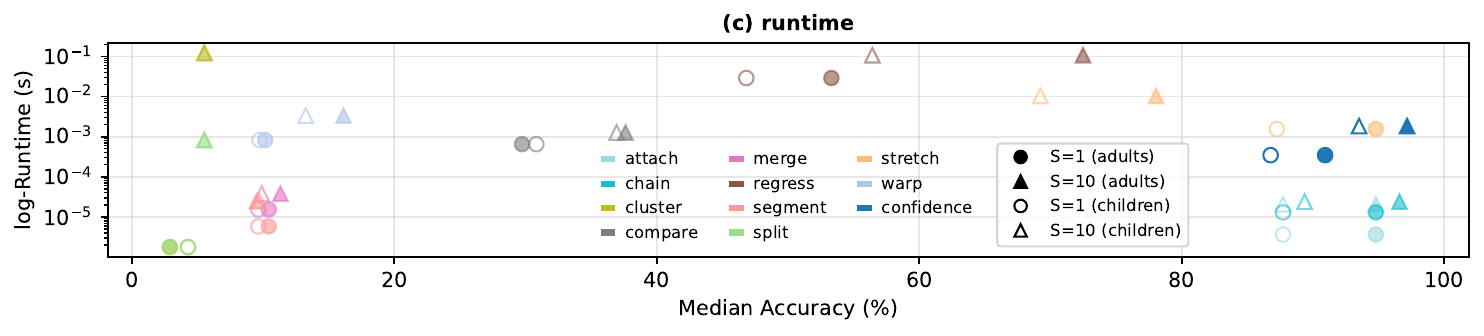}
  \caption{Top: Benchmark analysis of algorithms on median line assignment accuracies for (a) adult and (b) children reading gaze data. In post hoc analysis, the confidence algorithm performs within the benchmark. Its performance drops below (adults) or to (children) baseline accuracy (\texttt{attach}) for $S=1$ but then regains accuracy at (adults) or even above (children) baseline level. Bottom: Processing time vs. median accuracy for ad hoc analysis with $S=1, 10$ (c). CONF-LA shows best accuracy with slightly increased latency.}
  \label{fig:bm_seg_size_A_real_time_analysis}
  \Description{
  A three-part performance analysis. Top Plots (a and b) present two line graphs showing Median Accuracy against Segment Size for adults and children. Most baseline methods show a drop in performance from post hoc to ad hoc analysis. The "confidence" method (dark blue line) shows only slightly decreased performance for small segment sizes with sub optimal parameters. For optimal parameter settings, shown by red stars, accuracy remains high.
  Bottom Plot (c) is a scatter plot showing log-Runtime in seconds (y-axis) versus Median Accuracy (x-axis). Different shapes represent adults and children across segment sizes S=1 and S=10. The "confidence" markers (dark blue circles and triangles) are clustered at the far right of the x-axis, indicating high accuracy, with a runtime centered around one millisecond seconds. While slightly slower than some low-accuracy baselines, CONF-LA provides the best accuracy-to-latency trade-off, especially for children's data.}
\end{figure}

Figure \ref{fig:bm_seg_size_A_real_time_analysis}(c) compares runtime (milliseconds-per-segment on a standard laptop) and median accuracy across algorithms, to evaluate feasibility in real-time ($S=1$, circles) and ad hoc ($S=10$, triangles) assignment. 
CONF-LA requires slightly more time (1: 0.348~ms, 88~\% (children)/ 95 ~\% (adult); 10: 1.835~ms, 94.2 ~\%/97.7~\%) than absolute-positional baseline algorithms \texttt{attach} (1: 0.004~ms, 87.7~\%/94.8~\%; 10: 0.021~ms, 87.7~\%/94.8~\%) and \texttt{chain} (1: 0.013~ms, 87.7~\%/94.8~\%; 10: 0.024~ms, 89.4~\%/96.6~\%).   
This runtime is compatible with ad hoc deployment on standard CPUs.

For ad hoc assignment ($S=10$), CONF-LA consistently outperforms \texttt{attach} and \texttt{chain} across age groups, with particularly high gains for noisy children's data.
As shown in Figure \ref{fig:bm_seg_size_A_real_time_analysis}(a), this holds across all ad hoc segment sizes ($S>1$), indicating that modeling explicit noise and transition structures improves stability under increased signal variability, apparent in children's data.
For real-time assignment, where this temporal information is limited, CONF-LA's performance is slightly below baseline on adult data (see Figure \ref{fig:bm_seg_size_A_real_time_analysis}(b)), only reaching baseline performance with segment-specific parameter settings (red stars).
However, accuracy stabilizes with small segment sizes ($S\approx 5$), leveling or exceeding baseline performance.
These results suggest a trade-off inherent to using CONF-LA: 
Larger segment sizes yield processing delays, not only due to processing time but also buffering time until the number of required fixations is collected, while leveraging temporal information increases accuracy, especially under noisy conditions.

\subsection{Degradation Analysis}
To approximate webcam-like conditions, we introduced low-fidelity degradation to the Carr dataset.
While this does not replace evaluation on fully-labeled webcam datasets, it provides a stress test for sensitivity to combined spatial noise and lower time resolution.
As presented in Table \ref{tab:generalise_analysis}, both CONF-LA and the \texttt{attach} baseline algorithm are sensitive to degradation.  
However, the relative drop is smaller for CONF-LA, especially for segment sizes $S=10,N$, while pure filtering ($S=1$) does not yield meaningful improvement. 
This reinforces the improved resilience to spatial noise due to information integration across fixations.
In general, this analysis suggests that CONF-LA can improve ad hoc line assignment under degraded vertical accuracy, although performance remains dependent on distortion severity.

\subsection{Generalization Analysis}

\begin{table}
    \centering
    \caption{Median assignment accuracy of CONF-LA and the \texttt{attach} algorithm on different datasets using parameter settings optimized for the Carr dataset. Higher accuracy indicates that CONF-LA generalizes across datasets without individual parameter optimization. }
    \label{tab:generalise_analysis}
    \begin{tabular}{lcccccc}
        \toprule
        \textbf{dataset} & \multicolumn{2}{c}{Carr} & \multicolumn{2}{c}{Carr (low-fidelity)} & \multicolumn{2}{c}{MECOde} \\ 
        \textbf{Algorithm} & \texttt{attach} & \texttt{confidence} & \texttt{attach} & \texttt{confidence} & \texttt{attach} & \texttt{confidence} \\ 
        $S=N$ & 93.5\% & 96.6\% & 67.4\% & 78.5\% & 86.7\% & 90.3\% \\
        $S=10$ & 93.5\% & 97.8\% & 67.4\% & 78.5\% & 86.7\% & 91.5\% \\ 
        $S=1$ & 93.5\% & 95.6\% & 67.4\% & 68.8\% & 86.7\% & 88.5\% \\
        \bottomrule
    \end{tabular}
\end{table}

Our analysis so far, as well as the grid search for best parameters (see Appendix \ref{apdx:grid_search}), has used the annotated dataset provided by \citet{carr2022AlgorithmsAutomatedCorrection}. 
To test the generalization capabilities across different experimental setups and reading patterns, we evaluate CONF-LA on the MECOde dataset without parameter re-tuning.
Table \ref{tab:generalise_analysis} compares the resulting median accuracies. 
Across segment sizes, CONF-LA consistently outperforms the baseline, with the strongest gains for $S=10$, indicating the benefit for intermediate local smoothing.
Reading progression matrices (see Appendix \ref{apdx:behavioural_priors}) show that the transition matrix sufficiently captures reading progression structure allowing for generalization.
Overall, this analysis shows CONF-LA's capability to yield stable performance improvements across datasets without explicit training.

\subsection{Algorithm Analysis} \label{sec:A_analysis}

\begin{figure}
  \centering
  \includegraphics[width=\linewidth]{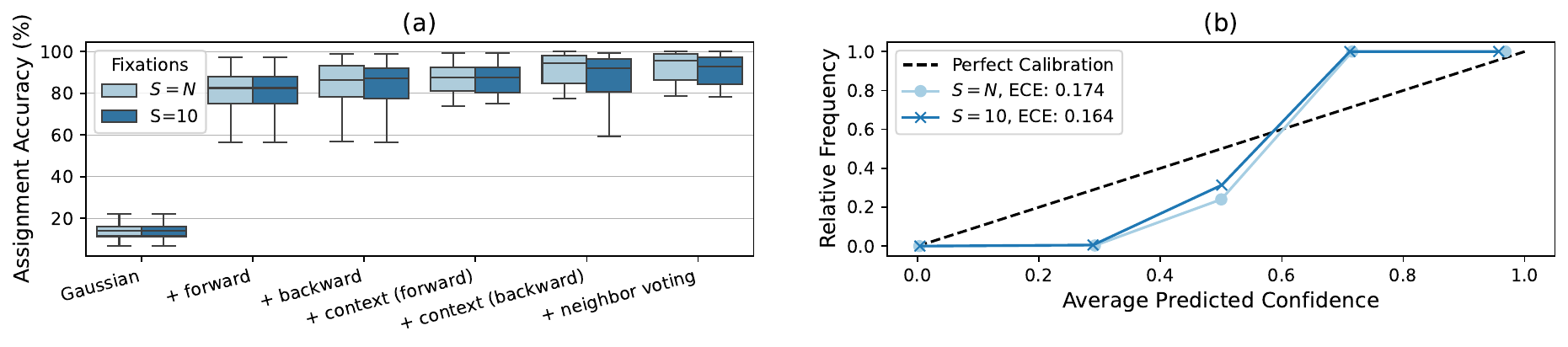}
  \caption{(a) Ablation analysis for post hoc and ad hoc assignment shows that integrating information over time and introducing context-sensitivity improve the CONF-LA's accuracy, respectively. (b) However, confidences are not yet optimally calibrated, which limits their reliability for use in downstream tasks.}
  \label{fig:alg_components}
  \Description{Two plots (a and b) analyzing model components and calibration. Sub-figure (a): A box plot of Assignment Accuracy showing the cumulative effect of adding model components. From left to right, accuracy increases significantly from the baseline "Gaussian" (approx. 15\%) 
  by adding a context-free forward pass, and then increases further step-wise with the addition of a context-free backward pass, then the context sensitivity, and finally peaking with the addition of the neighbor voting. Both segment sizes S=N and S=10 show similar upward trends. Sub-figure (b): A calibration plot showing Relative Frequency (y-axis) against Average Predicted Confidence (x-axis). A dashed diagonal line represents "Perfect Calibration." Both the S=N and S=10 lines stay below the diagonal for mid-range confidence (under-confident) and then cross over, indicating the model's confidence scores are not yet well calibrated.}
\end{figure}

Figure \ref{fig:alg_components}(a) illustrates the incremental contribution of each model component to assignment accuracies.
Using only the Gaussian line likelihood as a confidence score yields low accuracy. 
Filtering with a stationary, context-free transition matrix, boosted accuracy above 60~\%,  
demonstrating the value of information integration for correct assignment. 
The context-sensitive transition matrix update and the backward pass further improve accuracy, with additive effects. 
Updating confidence values through neighbor voting achieves the best accuracy, especially for ad hoc assignment. 

Confidence-scores could move beyond purely corrective purposes by introducing a model that produces not only more accurate, but also more informative outputs.
Rather than treating gaze-line assignment as a discrete decision, CONF-LA holds potential to provide a probabilistic description of fixation alignment, aiming for interpretable quantities that could inform both downstream processing and theoretical insight into reading behavior.
To explore this potential, we evaluate the confidence calibration. 
In its current implementation, CONF-LA yields a high Expected Calibration Error for both $S=N$ ($ECE=0.173$) and $S=10$ ($ECE=0.160$).
The reliability diagram (averaged across lines) presented in Figure \ref{fig:alg_components}(b) reflects this in a sigmoid shape, i.e., strong underconfidence for accuracies below 50~\% and overconfidence for higher accuracies. 
We suspect two main root causes for this miscalibration. 
First, we use a dataset that excludes non-reading fixations, resulting in a binary confidence of either reading a line or not. 
This approach is theoretically sound given its primary goal of enhancing accuracy in ad hoc line assignment, but it limits the algorithm's generalizability and downstream usability. 
Second, we use a hardcoded transition matrix, with default values updated only when context indicates a line change. 
These transition distributions are artificial and lack true uncertainty measures. 
Replacing them with parametrized distributions could improve calibration and enable downstream applicability.

\subsection{Analysis Summary}
While CONF-LA’s confidence-scores are not reliable enough for direct downstream use, they provide a probabilistic framework for interpreting reading behavior. 
The emission parameters reflect spatial precision and fixation variability, while the transition matrix encodes temporal reading patterns. 
Together, these components offer insights into individual reading dynamics beyond discrete line decisions.
Our analysis shows that CONF-LA already supports accurate ad hoc correction and generalizes across datasets without retraining, but still demands a trade-off against minimal delay. 

\section{Discussion}

\subsection{Probabilistic Modeling Perspective}

In the present work, we use a non-homogeneous HMM with Gaussian emissions and perform forward-backward inference using analytically computed, context-sensitive, and time-varying transition probabilities.
Conceptually, this model is related to the “Kalman + HMM” formulation \cite{bottos2019NovelSlipKalmanFilter, bottos2020TrackingProgressionReading}, yet it differs in that the continuous and discrete components are integrated rather than cascaded.
This allows us to perform continuous, ad hoc filtering across and local offline smoothing within sequential segments to infer confidence scores for line assignments.
However, the current confidence scores are internal model metrics rather than calibrated probabilities.
We have separated confidence-based line assignment from the HMM implementation for two reasons. 
First, the confidence measure itself is model-agnostic, i.e., it does not rely on an HMM for computation but can represent a likelihood value, a posterior probability, or a vote from multiple statistical models. 
Second, the defined HMM is only one possible probabilistic model capable of providing a line assignment posterior.
Nevertheless, this formalism can be extended to a fully parametrized model, in which parameters for transitions (e.g. Dirichlet) and emission (e.g. Normal-Inverse Gamma) are estimated from the data. 
This might improve calibration, also yielding more interpretable and reliable confidences for downstream modules in assistive technologies. 
Alternatively, adding parametrized innovation noise to the transition matrix would keep the generalizability benefits identified in this analysis while providing explicit transition uncertainty.
However, this must strike a balance with the runtime-accuracy tradeoff. 

\subsection{Contrasting CONF-LA to Learning-based Algorithms}
Importantly, inference in this framework is based on implied knowledge rather than data-driven learning.
While state-of-the-art post hoc methods use learning-based architectures  \citet{mercier2024DualInputStream, shangareev2025ReadingProgressTracking}, CONF-LA targets line assignment for assistive technologies, where data, computational resources, and latency are limiting constrains.
Our analysis shows that CONF-LA doesn't match reported post hoc accuracies (e.g. 99.63~\% on the Carr dataset in \citet{mercier2024DualInputStream}), but addresses limitations of these learning-based approaches.

First, learning-based methods, including \citet{bottos2019NovelSlipKalmanFilter, bottos2020TrackingProgressionReading}, require training on large datasets of labeled reading trials, whereas CONF-LA includes explicit behavioral assumptions and hence reduces the need for training to a hyper-parameter optimization to maximize performance. 
CONF-LA can generalize to unseen data of similar recording settings.
This is especially valuable in scenarios where labeled data is unavailable. 
Second, many SOTA learning-based methods typically require 
GPU training acceleration and may sustain higher inference cost.
CONF-LA, on the other hand, enables assignment with a delay of around 1~ms on standard CPUs, due to a lightweight inference procedure.
This aligns with the needs of assistive systems, where minimal latency is essential for usability. 
Third, CONF-LA's probabilistic framework does not require pre-processing of fixation information, such as normalization, and is invariant to font size and type, unlike e.g. \cite{mercier2024DualInputStream}. 
It is therefore setup agnostic and context independent.

Finally, CONF-LA offers the integration into real-world reading behavior, where reading can not only include jumps across more than one lines (explicitly excluded by \citet{shangareev2025ReadingProgressTracking}), but particularly fixations that do not represent reading behavior (explicitly excluded by \citet{mercier2024DualInputStream}). 
To the best of our knowledge, this is the first assignment algorithm that explicitly allows for fixations to stay unassigned, thus accounting for non-reading fixations. 
This analysis evaluated CONF-LA within the context of established algorithms for post-hoc assignments.
Thus, the analysis of non-reading fixations is beyond the scope of this paper, but is highly encouraged as a next step and should include another calibration analysis. 

\subsection{Extensions and Advancements on the Algorithm}
The current algorithm employs several fixed parameters inspired by typical skilled reading behavior. 
Although our analysis validated these choices for skilled readers, their fixed nature presents an opportunity for improvement. 
Specifically, future research should examine if and how well parameters can be adapted to capture different reading behaviors.  
Although not strictly required, adjusting parameters can provide insight into individual reading patterns.
Furthermore, CONF-LA overwrites context information rather than integrating it over time, thereby losing details about saccadic dynamics and introducing hand-tuned constants.
Similar to measurement noise, uncertainty in line transitions could be better modeled using suitable probability distributions.
This would allow information about gaze behavior to be integrated over time rather than replaced, potentially yielding more stable line-assignment estimates.
Once the uncertainty in gaze assignment is quantified probabilistically, model parameters themselves can become adaptive, reflecting individual reading styles or changing noise conditions.
Sampling from the posterior could enable the computation of prediction errors, which in turn could guide the adaptation of certain parameters.
In this sense, adaptivity would emerge as a by-product of full probabilistic modeling.
This adaptivity might especially improve parameter sensitivity in smaller segment sizes and, thus, stabilize performance. 

A preliminary example of adaptivity is already present in the mean-shift update, where likelihood distributions are slightly adjusted based on recent gaze behavior and resulting line assignment. 
Allowing for more incremental shift adaptation within a larger range (e.g., shift by average deviation with $\mu_{shift} \in [0.85, 1.15]$), could target CONF-LA's susceptibility to extreme slope and shift.
Alternatively, these limitations could also be addressed with an adaptive $\sigma_{scale}$ (e.g. allow for more variance in lower lines).
In general, adapting parameters is again another computational step that should be kept as light-weight as possible to ensure real-time capability of the correction. 

\subsection{Evaluation and Future Work}
Reading support systems can not only benefit from CONF-LA's real-time capability, but also leverage probabilistic information of gaze–text alignment with existing gaze-driven interaction paradigms.
Such integration would require a calibrated confidence and a more accurate differentiation between true reading behavior, i.e., the goal-directed processing of text, and non-reading gaze activity such as mind wandering or exploration \cite{rayner1996MindlessReadingRevisited}, which remains an open challenge across correction algorithms.
Algorithmic and probabilistic correction approaches, like CONF-LA, could be leveraged to address this challenge. However, such analysis requires appropriate accuracy measures and datasets that better capture the distinction between unassigned and assigned fixations.

Apart from this specific type of dataset, further evaluation of the CONF-LA approach on multiple datasets with lower resolution is required as this analysis uses synthetic degradation after fixation detection, which does not replace evaluation on real webcam data.
More variable text conditions should also be evaluated, including, but not limited to: 
(i) source-code reading \cite{almadi2024AdvancingDynamicTimeWarp, shamy2023IdentifyingLinesInterpreting}, 
(ii) non-Latin alphabetic systems, particularly vertical and right-to-left languages, which could be captured by adapting the behavioral prior, i.e., mirroring the transition matrix, and 
(iii) more dense and complex reading environments. 
We suggest exploring further application domains beyond reading where vertical noise is present in eye tracking data. 
Behavioral priors could be adapted to capture vertical gaze progression, e.g., scrolling through social media posts.
We also fitted the algorithm on the full dataset to demonstrate its potential, but strongly emphasize the need for broader studies that include cross-validation and real-time deployment.
Finally, integration into assistive technologies would enable real-world evaluation of technical performance (real-time capability) and user benefits (perceived and measurable improvements). 
We emphasize that such integration should always ensure privacy and accessibility, besides usability. 

\subsection{Advice for Practitioners}
Use default parameters with $S=10$ with our optimized hyperparameters (see Table \ref{tab:best_setting}). 
For real-time use, consider processing smoothing in overlapping fixation segments or consider fitting parameters to your specific setup. 
Adjust emission $\sigma$ to your eye-tracker’s noise level. 
Furthermore, allow for larger, incremental adaptation in $\mu_{shift}$ to capture extreme slope and drift patterns if expected.
For non-reading tasks (e.g. scrolling), adapt the transition matrix.

\section{Conclusion}
In this paper, we introduced CONF-LA, a confidence-score-based approach for gaze-line assignment in multi-line reading. 
Our analysis validated CONF-LA within a benchmark of post hoc algorithms and as particularly promising for ad hoc assignment required by assistive reading technologies. 

\section{Social Impact Statement} 

As reading is the fundamental basis for accessing knowledge, we see our research as a contribution to fostering educational equity. 
However, we recognize that detecting fine-grained reading patterns and, in extension, gaining insight into cognitive processes bears risks for misuse. 
Therefore, we encourage maximal privacy standards when working with gaze reading data, not only in research but also in assistive technologies that leverage this information.

\begin{acks}
We acknowledge the work of all authors and contributors of the benchmarking study in \citet{carr2022AlgorithmsAutomatedCorrection}. 
This work was made possible due to their effort and open research approach.
\end{acks}

\bibliographystyle{ACM-Reference-Format}
\bibliography{driftCorrection}

\appendix

\section{HMM equations}
\subsection{Generative model} \label{apdx:gen_mod}

This appendix provides a detailed mathematical specification of the HMM used in this work, which models $M$ lines as hidden states $X_i$ and vertical position as observations $Y_i$.
\begin{itemize}
    \item \textit{Prior distribution} is uniform over all $M$ lines \begin{math}
        P(X_1=l) = \frac{1}{M}
    \end{math}
    \item \textit{Transition distribution} is given by a $M\times M$ transition matrix: \begin{math}
        P(X_i^{(s)}=l | X_{i-1}^{(s)} = j) = L_i^{(s)}(j,l)
    \end{math}, where $L_i$ is a context-sensitive adaptation of the general transition prior encoded in $L$.
    \item \textit{Emission distribution} is normally distributed: \begin{math}
        P(Y_i^{(s)}=y|X_i^{(s)}=l) \sim \mathcal{N}(\mu_l, \sigma)
    \end{math}, where parameters are given by 
    \begin{itemize}
        \item shifted center of line $l$: \begin{math}
            \mu_l = y_l\,\cdot\,\mu_{shift}
        \end{math}, with adaptable $\mu_{shift} \in \mathbb{R}^+$ 
        \item scaled line distance $d_l$: \begin{math}
            \sigma = d_l \cdot \sigma_{scale}
        \end{math}, with fixed scaling factor $\sigma_{scale} \in \mathbb{R}^+$
    \end{itemize}
    \item \textit{Joint distribution} over $S$ fixations within a segment $s$:\begin{equation}
        P(X_{1:S}, Y_{1:S}) = P(Y_1^{(s)}|X_1^{(s)})\, P(X_1) \prod^S_{i=2} P(Y_i^{(s)}|X_i^{(s)})\,P(X_i^{(s)} | X_{i-1}^{(s)})
    \end{equation} 
\end{itemize}

\subsection{Message Passing Equations for Forward Backward Inference} \label{apdx:equations}

The inference procedure on the generative model defined in Section \ref{sec:gen_mod} is given by computing confidence scores through message passing.
We assume sequential processing of segments $s$ containing $S$ fixations each and formalize the forward pass as a continuous online filtering over these segments 
using the initial belief vector $\alpha_0^{(s=1)} = P(Y_1|X_1)\,P(X_1)$.
Following segments start with the final posterior belief from the previous segment $\alpha_0^{(s)} = \gamma_S^{(s-1)}$ for $s>1$. 
The backward pass, on the other hand, is formalized as a local, offline smoothing on a given segment of size $S>1$. 
The initial belief vector for the backward pass for all segments $\beta_S^{(s)}$ is an all-one vector of size M. 
The smoothed posterior within a segment is proportional to the product of the normalized belief messages $\alpha_i^{(s)}$ and $\beta_i^{(s)}$ in a current fixation $i$.
The final assignment posterior is obtained through normalization.
Detailed message passing formulas used for probabilistic inference are defined as follows: 
\begin{itemize}
    \item For every new fixation $i=1, ..., S$, we compute for all lines $l$
\begin{equation}
    \tilde{\alpha}_i^{(s)} (l) \propto d_i^{(s)}(l) \sum_{j=1}^M \mathbf{L}_i^{(s)}(j,l){\tilde\alpha}_{i-1}^{(s)}(j)
    \ \ \ \ \ \ \ \ \ \ \ \ \ \ \ \ \ \ \ \ 
    \alpha_i^{(s)} (l) = \frac{\tilde{\alpha}_i^{(s)} (l) }{\sum_{l'} \tilde{\alpha}_i^{(s)} (l') }
\end{equation}

    \item Recursively for fixations $i = S-1, ..., 1$, we compute for all lines $l$
\begin{equation}
    \tilde{\beta}_i^{(s)} (l) \propto \sum_{j=1}^M \mathbf{L}_{i+1}^{(s)}(l,j)d_{i+1}^{(s)}(j)\tilde{\beta}_{i+1}^{(s)}(j) 
    \ \ \ \ \ \ \ \ \ \ \ \ \ \ \ \ \ \ \ \ 
    \beta_i^{(s)} (l) = \frac{\tilde{\beta}_i^{(s)} (l) }{\sum_{l'} \tilde{\beta}_i^{(s)} (l') }
\end{equation}

    \item Finally, we combine messages into a posterior:    
\begin{equation}
    \gamma_i^{(s)}(l) \propto \alpha_i^{(s)}(l) \beta_i^{(s)}(l)
    \ \ \ \ \ \ \ \ \ \ \ \ \ \ \ \ \ \ \ \ 
    \gamma_i^{(s)}(l) = \frac{\alpha_i^{(s)}(l) \beta_i^{(s)}(l) }{\sum_{l'} \alpha_i^{(s)}(l') \beta_i^{(s)}(l')}
\end{equation}
\end{itemize}
Finally, the filtered belief $\alpha_S^{(s)} = \gamma_S^{(s)}$ is passed on to the next segment as $\alpha_0^{(s+1)}$ under the Markov property of the system.
It is worth noting the following: For ad hoc line assignment ($S=1$), inference becomes a mere filtering procedure. 
For post hoc line assignment ($S=N$), inference becomes a full smoothing procedure. 

\section{Analysis of behavioral prior} \label{apdx:behavioural_priors}
\begin{figure}[H]
  \centering
  \includegraphics[width=\linewidth]{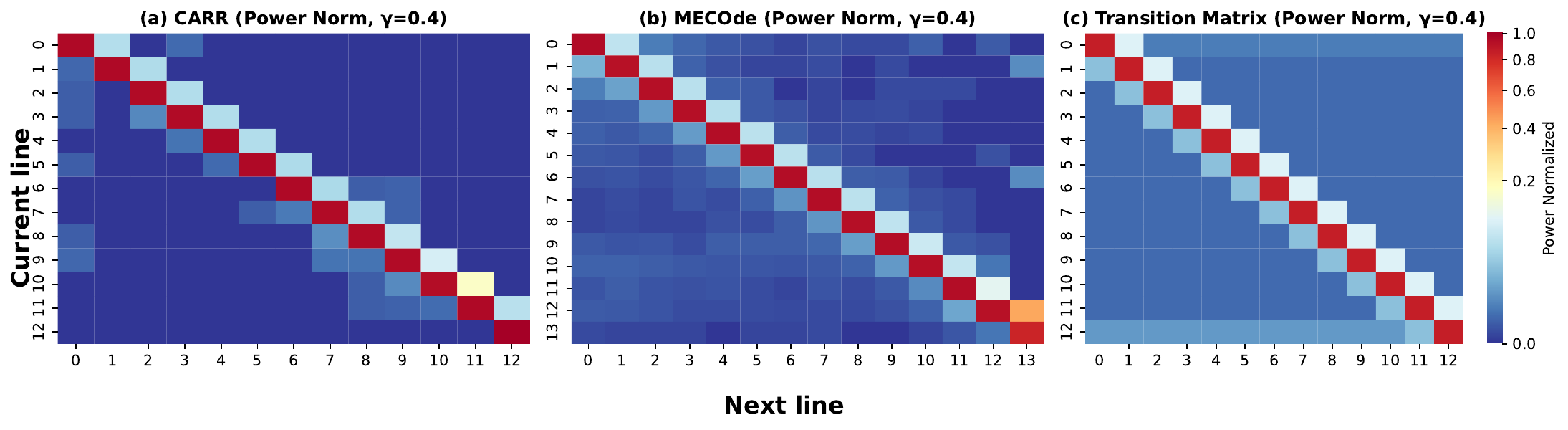}
  \caption{Reading progression matrices of the used data sets (a, b) are consistent with handcrafted priors in the transition matrix (c).}
  \label{fig:transition matrices}
  \Description{Three matrices with similar structures: The diagonal is red indicating the highest values. Two matrices show the average reading progression matrices from the Carr and MECOde datasets, the third one shows the handcrafted transition priors.}
\end{figure}

The chosen behavioral prior distribution for the transition matrix was crafted to resemble a discrete gamma distribution, reflecting the temporal dynamics of skilled reading (e.g., high probability of staying in the same line, lower probability of regressions or large jumps). 
While not directly learned from the data, Figure \ref{fig:transition matrices} shows that this design aligns with empirical reading progression patterns observed in datasets like CARR \cite{carr2022AlgorithmsAutomatedCorrection} and MECOde \cite{siegelman2022ExpandingHorizonsCrosslinguistic}. 
This verifies qualitative alignment with our hardcoded transitions, although we explicitly assign a slightly higher probability to jumps over more than one line.

\section{Research Methods}
\subsection{Grid Search} \label{apdx:grid_search}

\begin{figure}[ht]
    \centering
    \includegraphics[width=0.75\linewidth]{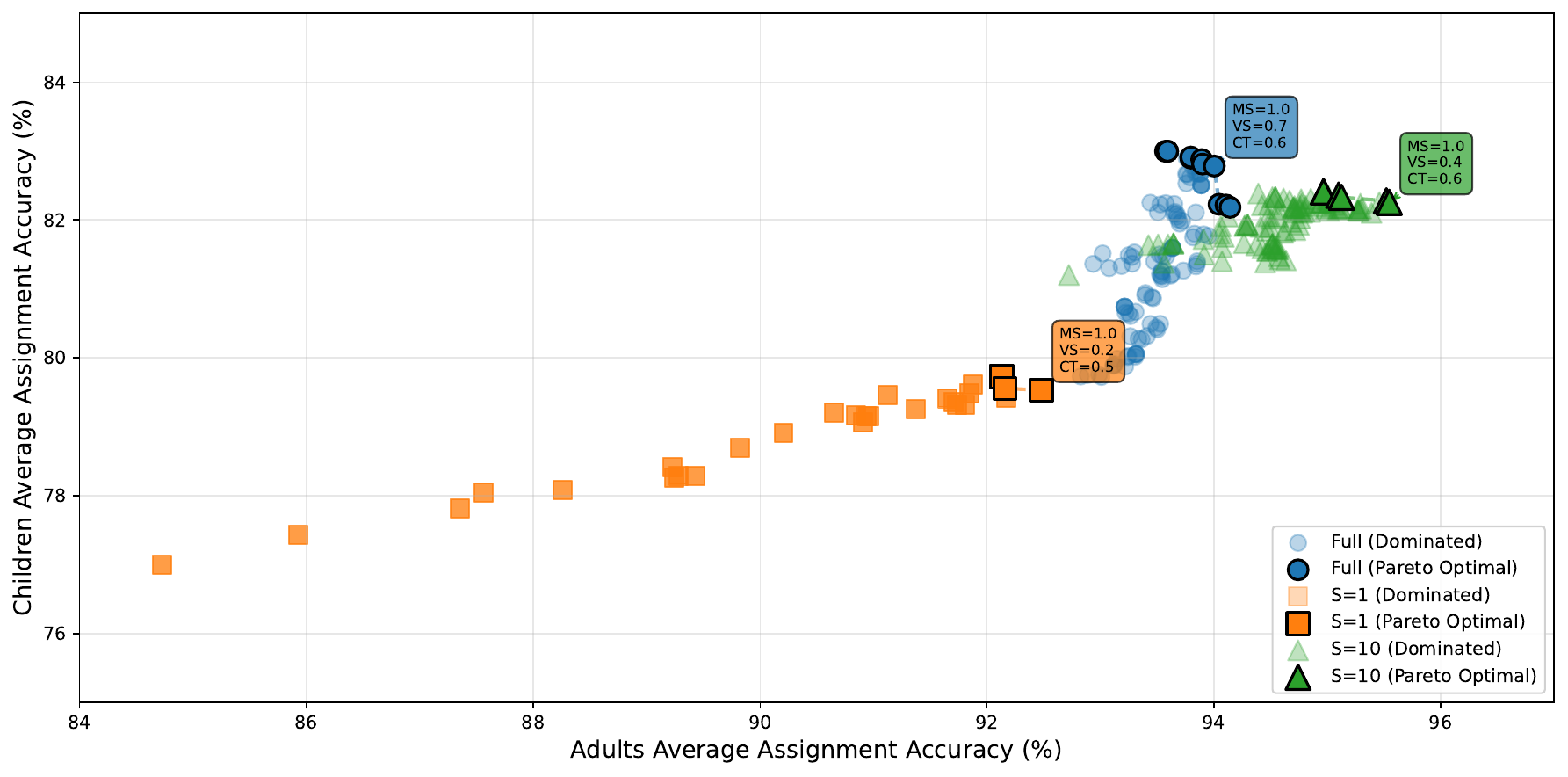}
    \caption{Parteo Analysis of grid search results.}
    \label{fig:parteo_frontier}
    \Description{A parteo plot showing grid search accuracies for children (y-axis, 76\% - 84\%) and adults (x-axis, 84\% - 96\%) for different segment sizes (1, 10, and N). Accuracies for segment size 1 are scattered along a line in the lower left corner of the plot. Segment sizes 10 and N are clustered in the upper right corner at higher accuracies. Segment size 10 shows more variance for children data, and segment size N shows more variance for adult data.}
\end{figure}

We conducted a grid search to find good-performance parameter settings and analyze the effect of different parameters in a parameter analysis. 
Table \ref{tab:grid_ranges} details the parameter space that was used in each search, respectively. 
Figure \ref{fig:parteo_frontier} visualizes grid search results by performance on children vs. adult reading data. 
Table \ref{tab:best_setting} presents best performing parameters across segment sizes.

\begin{table}[ht]
  \caption{Parameter ranges for grid search and parameter analysis.}
  \label{tab:grid_ranges}
  \begin{tabular}{cll}
    \toprule
    Parameter & Grid search & Parameter analysis \\
    \midrule
    $S$ & [1, 10, N] & [1, 10, N] \\
    $\mu_{shift}$ & [0.97, 0.98, 0.99, 1.0, 1.01, 1.02] & [0.98, 0.99, 1.0, 1.01, 1.02] \\
    $\sigma_{scale}$ &  [0.4, 0.5, 0.6, 0.7, 0.8] & [0.2, 0.4, 0.6, 0.8, 1] \\
    $C$ & [0.5, 0.6, 0.7, 0.8] & [0.1, 0.2, 0.3, 0.4, 0.5, 0.6, 0.7, 0.8, 0.9, 1] \\
    $\Omega$ & [0.5, 0.6, 0.7, 0.8] & [0.6] \\
  \bottomrule
\end{tabular}
\end{table}

\begin{table}
  \caption{Parameter combinations from grid search for best combined line assignment accuracy averaged over all reading trials. Distinct accuracies on adult and children data show a performance gap between these two groups. Multiple values for one parameters indicate identical best accuracy values.}
  \label{tab:best_setting}
\begin{tabular}{rccccccc}
    \toprule
     $S$ & $\mu_{shift}$ & $\sigma_{scale}$ & $C$ & $\Omega$ & \multicolumn{3}{c}{Mean Accuracy (in \%)} \\
         &               &                  &     &          & \textit{Adults} & \textit{Children} & \textit{Combined} \\
    \midrule
     $N$ & 1.0 & 0.7 & 0.6 & 0.6 & 94 & 82.78 & 88.4\\
     1   & 1.0 & 0.2 & 0.5 & 0.5, 0.6 & 92.48 & 79.53 &  86\\
     10  & 1.0 & 0.4 & 0.6 & 0.6 & 95.55 & 82.25 & 88.9\\
     \midrule
     1 & 1.0 & 0.2 & 0.5 & 0.5,0.6,0.7,0.8 & 92.48 & - & -\\
     2 & 1.0 & 0.2 & 0.7 & 0.6,0.7,0.8 & 93.84 & - & -\\
     3 & 1.0 & 0.4 & 0.7 & 0.7,0.8 & 94.65 & - & -\\
     4 & 1.0 & 0.4 & 0.6 & 0.8 & 94.48 & - & -\\
     5 & 1.0 & 0.4 & 0.6 & 0.7,0.8 & 94.72 & - & -\\
     \midrule
     1 & 1.0 & 0.3 & 0.6 & 0.5,0.6,0.7,0.8 & - & 79.73 & -\\
     2 & 1.0 & 0.2 & 0.5 & 0.5,0.6,0.7,0.8 & - & 81.32 & -\\
     3 & 1.0 & 0.5 & 0.6 & 0.7,0.8 & - & 81.63 & -\\
     4 & 1.0 & 0.6 & 0.5 & 0.6,0.7,0.8 & - & 82.34 & -\\
     5 & 1.0 & 0.6 & 0.6 & 0.5,0.6 & - & 82.56 & -\\

  \bottomrule
\end{tabular}
\end{table}

\subsection{Degradation Procedure} \label{apdx:degrad}
The following algorithm implements the selective decimation used to simulate low-fidelity data.\\
\textbf{Algorithm: Temporal Binning with Noise Injection Input}\\
\textbf{Input:} \texttt{fixations} (high-freq), \texttt{bin\_dur} (ms), \texttt{params} (noise, shift, slope)\\
\textbf{Output:} \texttt{degraded\_fixations}
\begin{enumerate}
    \item Initialize \texttt{last\_end = 0}, \texttt{first\_x = fixations[0].x}
    \item For each f in fixations:
    \begin{enumerate}
        \item If \texttt{(f.start - last\_end) < bin\_dur}: Skip (simulate lower clock rate)
        \item else:
            \begin{itemize}
                \item Align Timestamps:\\
                    \texttt{d\_start = Ceiling(f.start / bin\_dur) * bin\_dur}\\
                    \texttt{d\_end = Floor(f.end / bin\_dur) * bin\_dur}
                \item Apply Degradation according to Eq. \ref{eq:y_distortion_sim} if \texttt{d\_start < d\_end} and if \texttt{d\_end - d\_start > bin\_dur} (drop fixations that would not have been captured by low-frequency device)
                \item Store new fixation data in \texttt{degraded\_fixations}
                \item Update \texttt{last\_end = d\_end}
            \end{itemize}
    \end{enumerate}
    \item Return \texttt{degraded\_fixations}
\end{enumerate}

\subsection{Use of LLMs}
During the completion of this work, LLMs were used to assist with coding, optimizing the argumentative flow, and providing feedback on some aspects of the work (e.g., writing style or grammar).

\section{Example gaze path correction}
\begin{figure}
  \centering
  \includegraphics[width=0.9\linewidth]{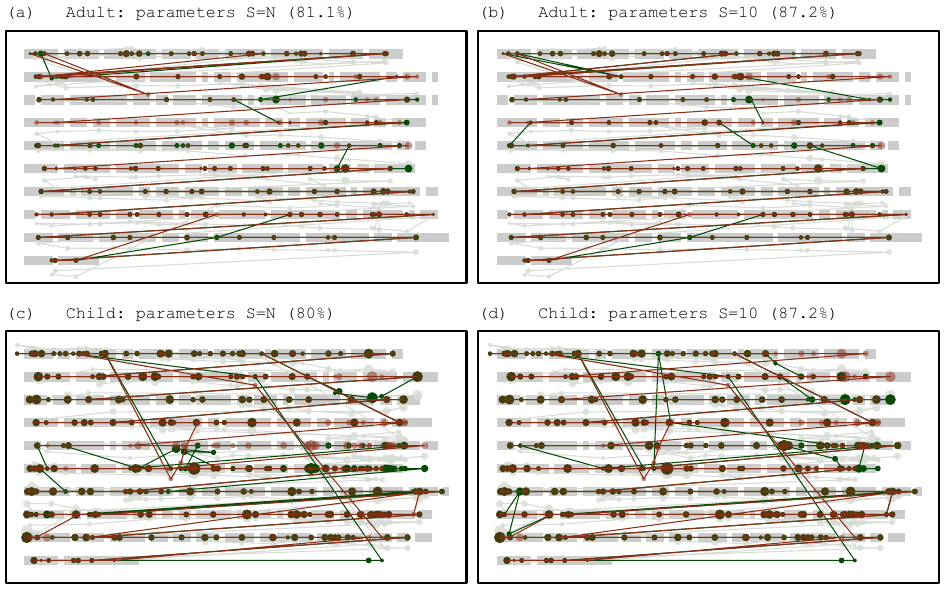}
  \caption{Corrected gaze paths for post hoc (left) and ad hoc (right) correction of selected reading trials from the \citet{carr2022AlgorithmsAutomatedCorrection} dataset. Eye movements were recorded with EyeLink 1000 at sampling rate of 1000~Hz. Uncorrected fixations (grey) are corrected by our CONF-LA approach (green) and compared to a manual gold correction (red) to compute line assignment accuracies. }
  \label{fig:Eyelink_path}
  \Description{Four plots showing a different reading gaze path each for adults (upper row) and children (lower row). For each age group, the left column shows a bad correction example, and the right column shows a good correction example.}
\end{figure}

\begin{figure}
  \centering
  \includegraphics[width=\linewidth]{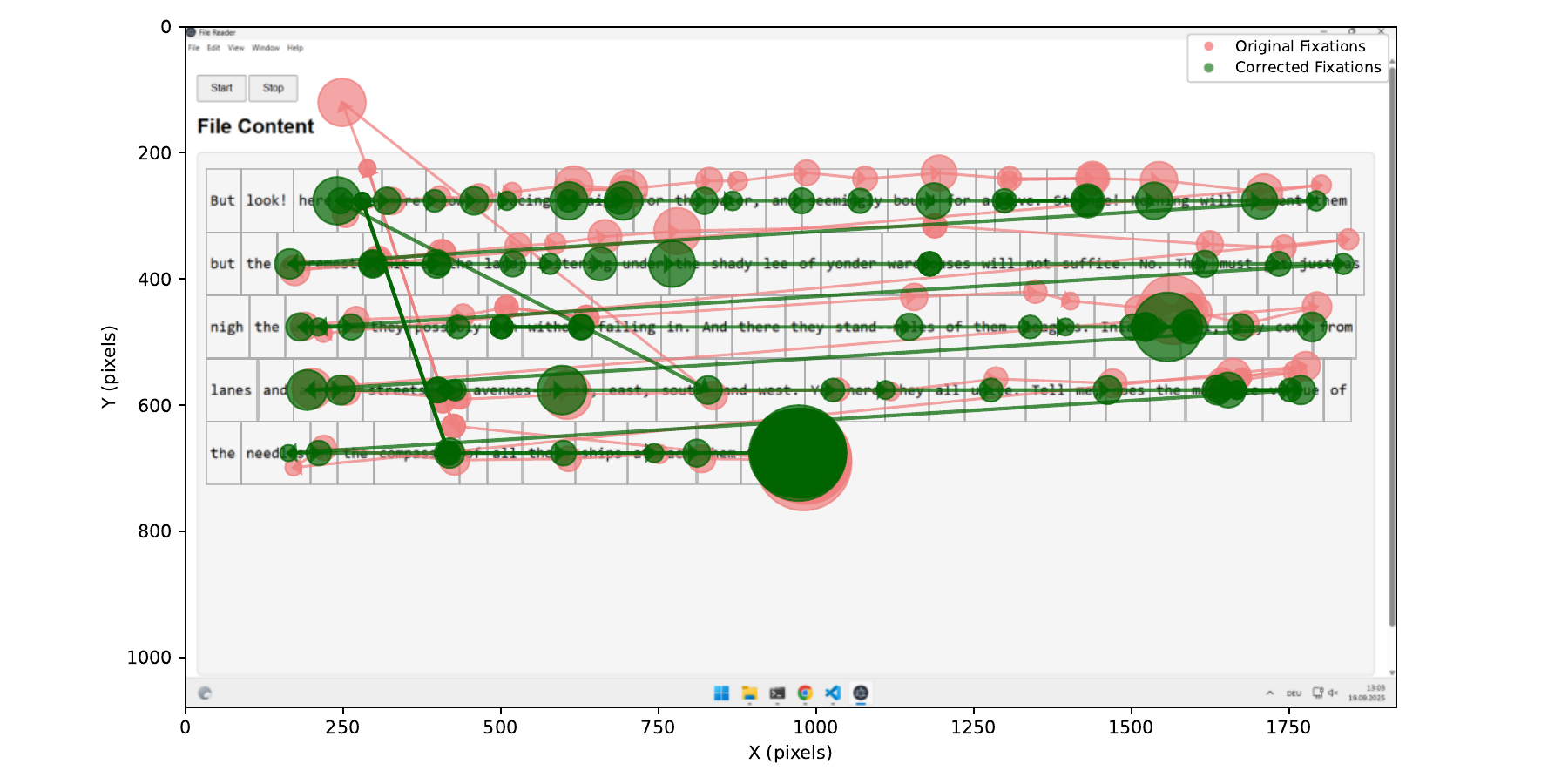}
  \caption{Ad hoc correction of eye movements that were recorded with Tobii Pro Fusion at a sampling rate of 250~Hz. Fixations were computed using IV-T. Segments contained \~2 seconds of eye-tracking data, yielding 3-12 fixations per segment.   
  Uncorrected fixations (pink) are corrected by our CONF-LA approach (green). No manual gold correction exist for this data.}
  \label{fig:Tobii_path}
  \Description{An example gaze path that has been corrected ad hoc. The original gaze path is shown in pink, the corrected gaze path is shown in green.}
\end{figure}

Figure \ref{fig:Eyelink_path} compares a correction with lower (left column) and higher (right column) mean accuracies for adults (top) and children (bottom), respectively.
These gaze path examples are taken from the data set by \cite{carr2022AlgorithmsAutomatedCorrection} which was recorded with a sampling rate of 1000~Hz. 
Figure \ref{fig:Tobii_path} presents a gaze path example recorded at a sampling rate of 250~Hz without annotation. 
We encourage the creation of a low-fidelity, annotated reading gaze data set for future analysis.

\end{document}